\documentclass{ws-procs9x6}

\begin{document}

\title{Energy Dependence of Helicity--Flip Amplitudes in the Small--Angle
 Elastic pp--Scattering}

\author{S.~M. TROSHIN and N.~E. TYURIN}

\address{Institute
for High Energy Physics,\\
Protvino, Moscow Region, 142281 Russia}

\maketitle

\abstracts{We discuss a role of the double helicity-flip amplitudes and
derive  new unitary bounds for these
 amplitudes in
 elastic $pp$-scattering at small
values of $t$. We show that the usual assumption on the smallness of such amplidudes
can be justified only in the shrinking with energy region of small $t$ values.
}
Discussion of a role and magnitude of helicity-flip amplitudes in small-angle
elastic scattering
has a long history  and is an important issue in
the studies of the spin properties of diffraction. Recently
an interest in  accounting the contributions
of helicity-flip amplitudes becomes associated with
CNI polarimetry related problems\cite{brav} as well.
Not only non-flip and single helicity-flip amplitudes can give
contributions and affect  the estimates and bounds for the analyzing
power $A_N$. Double helicity-flip amplitudes can also contribute
into $A_N$ and  their behavior at high energies is also
 important  for the spin correlation
parameters and total cross-section differences in experiments with two
polarized beams available at RHIC nowadays.

The double helicity-flip amplitudes are usually
 neglected since they are supposed to be small
 in the whole region of momentum transfers. But this assumption  is based
merely on the technical simplification of the problem and is not
valid at large momentum transfers in elastic $pp$-scattering
where double-flip amplitudes can play an important role and fill up
multiple-dip structure in differential cross-section providing
correct description of the experimental data\cite{uf24}.
It is natural then to asses the role of double helicity-flip amplitudes
at small and moderate values of $t$ also.

The method we use is based on the
unitarity equation for helicity amplitudes of elastic
$pp$-scattering, i.e. we adhere to a
rational form of unitarization which
corresponds to an approximate wave function which changes both
the phase and amplitude of the wave in potential scattering.
 For the helicity amplitudes
 of $pp$--scattering (i.e. for the two--fermion
scattering)
 the corresponding solution of the unitarity equations :
\begin{eqnarray}
 F_{\lambda_3,\lambda_4,\lambda_1,\lambda_2}({\bf p},{\bf q}) & =
  & U_{\lambda_3,\lambda_4,\lambda_1,\lambda_2}({\bf p},{\bf q})+ \label{heq}\\
 & & i\frac{\pi}{8}
\sum_{\lambda ',\lambda ''}\int d\Omega_{{ \bf \hat k}}
U_{\lambda_3,\lambda_4,\lambda ',\lambda ''}({\bf p},{\bf k})
F_{\lambda ',\lambda '',\lambda_1,\lambda_2}({\bf k},{\bf q}),\nonumber
\end{eqnarray}
in the impact parameter representation
can be found an
explicit solution in the following
form:
\begin{eqnarray}
f_1 & = & \frac{(u_1 + u_1^2 - u_2^2)(1 + u_3 + u_4) - 2(1 + 2u_1 - 2u_2)u_5^2}
{(1 + u_1 - u_2)[(1 + u_1 + u_2)(1 + u_3 + u_4)- 4u_5^2]},\nonumber\\
f_2 & = & \frac{u_2(1 + u_3 + u_4) - 2u_5^2}
{(1 + u_1 - u_2)[(1 + u_1 + u_2)(1 + u_3 + u_4)- 4u_5^2]},\nonumber\\
f_3 & = & \frac{(u_3 + u_3^2 - u_4^2)(1 + u_1 + u_2) - 2(1 + 2u_3 - 2u_4)u_5^2}
{(1 + u_3 - u_4)[(1 + u_1 + u_2)(1 + u_3 + u_4)- 4u_5^2]},\nonumber\\
f_4 & = & \frac{u_4(1 + u_1 + u_2) - 2u_5^2}
{(1 + u_3 - u_4)[(1 + u_1 + u_2)(1 + u_3 + u_4)- 4u_5^2]},\nonumber\\
f_5 & = & \frac{u_5}
{(1 + u_1 + u_2)(1 + u_3 + u_4)- 4u_5^2},\label{fi}
\end{eqnarray}
where for simplicity we omitted in the functions $f_i(s,b)$
and $u_i(s,b)$ their arguments. Unitarity requires that
$\mbox{Re}u_{1,3}(s,b)\geq 0$, but the absolute values of the
functions $u_i(s,b)$ should not be limited by unity.
For the functions $u_{2,4}(s,b)$ we adhere to a simple general
 (using arguments
 based on the analytical properties in the complex $t$--plane\cite{f4n}):
\begin{equation}
u_{2}\sim u_4 \sim s^{\Delta} e^{-\mu b}.\label{usbn}
\end{equation}
To get an upper bound for the amplitudes $F_{2,4}(s,t)$
we consider the case when $u_{2,4}(s,b)$ are dominating ones.
Then we have for the amplitudes $F_{2,4}(s,t)$ the following
representation
\begin{equation}
F_{2}(s,t)=\frac{is}{\pi^2}\int_0^\infty bdb \frac{u_{2}(s,b)}
{1-u_{2}^2(s,b)}J_0(b\sqrt{-t})\label{f2s}
\end{equation}
and
\begin{equation}
F_{4}(s,t)=\frac{is}{\pi^2}\int_0^\infty bdb \frac{u_{4}(s,b)}
{1-u_{4}^2(s,b)}J_2(b\sqrt{-t})\label{f4s}
\end{equation}
Using for $u_{2,4}(s,b)$ the functional dependence in the form of Eq.
(\ref{usbn}) it can be  shown that  the amplitude $F_2(s,t=0)$
cannot rise faster than $s\ln s$ at $s\to\infty$ and the function
\[
\hat F_4(s,t=0)\equiv [\frac{m^2}{-t}F_4(s,t)]|_{t=0}
\]
cannot rise faster than $s\ln^3 s$ at $s\to\infty$.

Thus, we can state that the explicit account of unitarity in the
form of $U$ - matrix approach
leads to the following upper bound for the cross-section difference
\[
\Delta\sigma_T \leq c\ln s,
\]
where
\[
\Delta\sigma_T\equiv\sigma_{tot}(\uparrow\downarrow )-\sigma_{tot}(\uparrow\uparrow )
\sim-\frac{1}{s}\mbox{Im}F_2(s,t=0).
\]
It should be noted that the asymptotic behaviour of the amplitudes
$F_1$ and $F_3$ are determined by the functions $u_2$ and $u_4$, respectively, in the
situation when these functions dominate;  the
Froissart--Martin asymptotical bound  for these amplitudes remains under these
circumstances,  i.e. they are
limited by $cs\ln^2s$ at $t=0$.

Another related important  consequence is the conclusion on the possibility to neglect
helicity-flip amplitudes $F_2$, $F_4$ and $F_5$ under calculations of differential
cross-section
\[
\frac{d\sigma}{dt}=\frac{2\pi^5}{s^2}(|F_1(s,t)|^2+|F_2(s,t)|^2+|F_3(s,t)|^2+
|F_4(s,t)|^2+4|F_5(s,t)|^2)
\]
and double helicity-flip amplitudes $F_2$ and $F_4$ under calculation
of analyzing power $A_N$
\[
A_N(s,t)\frac{d\sigma}{dt}=\frac{2\pi^5}{s^2}\mbox{Im}[(F_1(s,t)+F_2(s,t)+F_3(s,t)
-F_4(s,t))^*F_5(s,t)]
\]
in the region of small values of $t$ in high energy limit. This conclusion is based
on the above bounds for the helicity amplitudes and their small $t$ dependence due to
angular momentum conservation, i.e.
at $-t\to 0$: $F_i \sim \mbox{const}$, $(i=1,2,3)$, $F_5\sim\sqrt{-t}$ and
$F_4\sim -t$. However, the dominance of the helicity-non-flip amplitudes ceases
to be valid at fixed values of momentum transfers, where , e.g. amplitude $F_4$ can
become a dominant one, since its energy growth is limited by the function $s\ln^3 s$, while
other helicity amplitudes cannot increase faster than $s\ln^2 s$.

One should recall that unitarity for the helicity amplitudes  leads to    a peripheral
dependence of the amplitudes $f_i(s,b)$ $(i=2,4,5)$
on the impact parameter $b$ at high energy, i.e.
\[
|f_i(s,b=0)|\rightarrow 0
\]
at $s\rightarrow\infty$. This is a consequence of the explicit
 unitarity
representation for the helicity amplitudes through the  $U$-matrix and it is this
fact allows
one to get better bounds for the helicity-flip amplitudes.

 Thus,  we have shown here and in\cite{uf5}, that
the following asymptotic results should be valid:
\begin{itemize}
\item
 the ratio
$r_5(s,0)\equiv2\hat{F}_5(s,0)/[F_1(s,0)+F_3(s,0)]$
cannot increase with energy,
\item
the amplitude $F_2(s,t=0)$
cannot increase faster than $s\ln s$,
\item
 the function
$\hat F_4(s,t=0)$
should not rise faster than $s\ln^3 s$ at high energies.
\end{itemize}
Nowadays
RHIC spin program includes experiments with two polarized proton beams at the
highest available energies and the above bounds could be useful
and provide grounds for the estimations of the spin observables
in the forward region in these experiments.
 The above bounds provide justification of
the smallness of the double helicity-flip amplitudes in the low-$t$ region, but
simultaneously they  imply an importance of the double helicity-flip
amplitudes at the moderate values of momentum transfers. This result is in accordance
with early analysis of experimental data performed in\cite{uf24}.
It is also evident that the region of the momentum transfers where helicity--flip
 amplitudes can be neglected is shrinking with energy, e.g. for the amplitude
$F_4$ this shrinkage is proportional to $1/\ln s$.
Magnitude of the helicity amplitude $F_2$ at $t=0$ can be measured directly
at RHIC through the measurements of $\Delta\sigma_T$\cite{brav}
 and it is definitely  an
important study of the spin properties of diffraction.
The experimental data for
$\Delta\sigma _T(s)$
could also be a useful source of information
on the low-$x$ behaviour of the spin structure function $h_1(x)$.
\section*{Acknowledgments}
We are grateful to the Local Organizing Committee of SPIN2004 for the warm hospitality
in Trieste during the Symposium.

\end{document}